# Proton Improvement Plan II Cryogenic Distribution System thermodynamic design

Ashish Kumar Shukla[1], Andrew Dalesandro[2], Ram Dhuley[2], and William Soyars[2]

[1]Raja Ramanna Centre for Advanced Technology, Indore, Madhya Pradesh, India
[2]Fermi National Accelerator Laboratory, PO Box 500, Batavia, Illinois, USA
e-mail: ashish.igcar@gmail.com

**Abstract**. The Proton Improvement Plan – II (PIP-II) is a superconducting linear accelerator being built at Fermilab that will provide 800 MeV proton beams for neutrino production. The Linac requires cooling at 40 K, 5 K, and 2 K temperatures, which will be provided by cryogenic helium produced by a Helium Cryoplant and distributed by a Cryogenic Distribution System (CDS). Based primarily on the Linac heat load requirements at each temperature and the allowable pressure drop, we have made a preliminary thermodynamic design of the CDS. The design also incorporates special requirements such as controlled and/or fast cooldown of the superconducting RF cavities and their dual maximum allowable working pressures. This paper presents the overall features of the PIP-II CDS, sizing of helium process circuits, different operating modes, and calculated mass flow capacities that cater to these operating modes.

## 1. Introduction

The PIP-II Linac contains a total of twenty-three cryomodules of five different types, each with a unique cavity design and heat load contribution, and three sets of RF frequencies. There is a half wave resonator (HWR) operating at 162.5 MHz, two types of single spoke resonators (SSR1, SSR2) at 325 MHz, and two types of elliptical cavities low beta and high beta (LB650, HB650) at 650 MHz. The PIP-II Cryogenic Distribution System (CDS) consists of one Distribution Valve Box (DVB), ~265 m of cryogenic transfer line, twenty-five modular Bayonet Cans (BCs) and one Turnaround Can (TC). Each cryomodule requires cooling at 40 K, 5 K, and 2 K temperatures that will be provided using cryogenic helium supplied from a Cryogenic Plant (CP) to the cryomodule Linac via the CDS.

## 2. CDS main features

Figure 1 shows a simplified schematic of the CDS, composed of a Distribution Valve Box (DVB) downstream of the CP interface, an Intermediate Transfer Line (ITL), and a Tunnel Transfer Line (TTL) located in the Linac tunnel including a Turnaround Can (TC). The CDS main process lines start at the DVB and span the entire CDS length to the TC, where the flow direction is reverted back to the DVB. The CDS TTL exchanges cryogenic helium with cryomodule loads through removable U-tubes (called 'branch lines'). Flows through the main process lines are in series while flows through the branch process lines are in parallel, as illustrated in Figure 2. In addition to cold process lines, the CDS also provides warm helium supply for controlled cooldown of the Linac as well as pressure relief vent lines for the cryomodules and the CDS itself. The CDS process circuits are named as follows:

1. 4.5 K Supply →
   - 4.5 K Supply main line;
   - 4.5 K Supply branch line;
2. 2 K Return →
   - 2 K Return main line;
   - 2 K Return branch line;
3. LTTS Return →
   - Low Temperature Thermal Shield Return main line;
   - Low Temperature Thermal Shield Return branch line;
4. HTTS Supply →
   - High Temperature Thermal Shield Supply main line;
   - High Temperature Thermal Shield Supply branch line;
5. HTTS Return →
   - High Temperature Thermal Shield Return main line;
   - High Temperature Thermal Shield Return branch line.
6. CD Return →
   - Cool Down Return main line;
   - Cool Down Return from HTTS Return
   - Cool Down Return from 2 K Return
   - Cool Down Return from LTTS Return

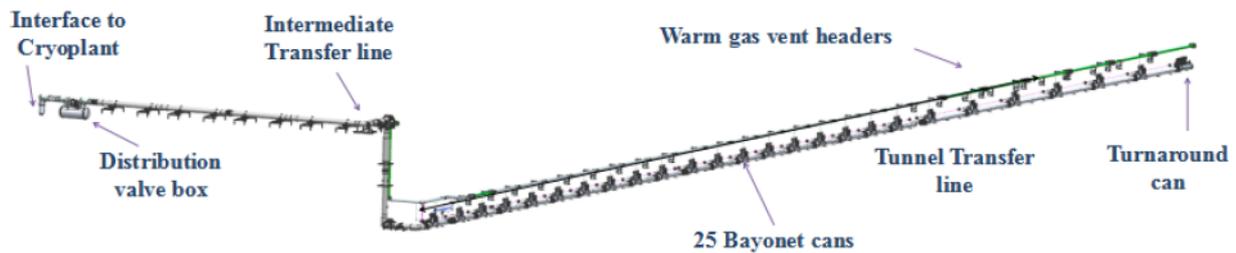

**Figure 1.** Schematic of the PIP-II Cryogenic Distribution System.

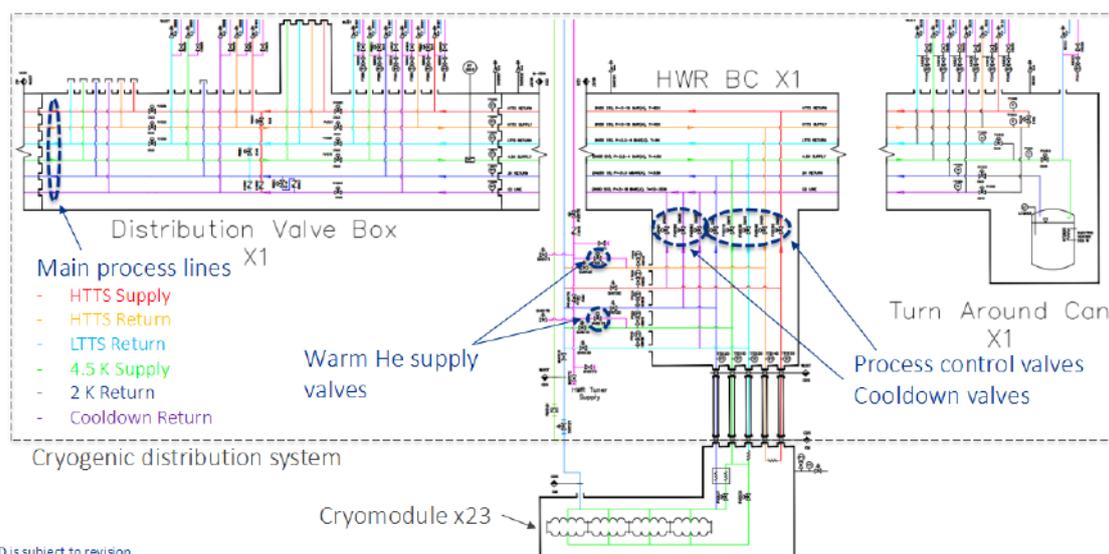

**Figure 2.** PID of the PIP-II Cryogenic Distribution System.

## 3. Design of the CDS for steady state Linac operation

Thermodynamic design of the PIP-II CDS is an iterative process [1-6]. We first assume process line and control valve sizes and calculate the expected heat load on the CDS process lines. This calculated CDS heat load is combined with the specified Linac heat load. For this total heat load we then calculate the required mass flow rate in each CDS circuit. The required mass flow rates are then compared with the calculated CDS mass flow rates capacity for pressure drop budget dictated by the Cryoplant. If the required mass flow rates are less than CDS mass flow capacity, then the configuration is selected. Otherwise, the above process is repeated using larger line sizes. Selected CDS process line details including diameter, length, and control valve size for the main and branch lines are given in Table 1.

Table 1. CDS Process Line Details

| Type of process | Process Line | Temp. [K] | Pipe Size | Length [m] | DVB Valves | Tunnel Valves |
|---|---|---|---|---|---|---|
| Main process lines | 2 K Return | 3.8 | DN250 | 250 | - | DN25, TC |
| | 4.5 K Supply | 4.5 | DN50 | 250 | 1 x DN40 | DN25, TC |
| | LTTS Return | 9 | DN50 | 250 | 2 x DN40 | DN25, TC |
| | HTTS Supply | 40 | DN50 | 250 | 2 x DN20 | - |
| | HTTS Return | 80 | DN50 | 250 | 2 x DN20 | DN15, TC |
| | CD Return | 80 | DN80 | 250 | 1 x DN65 | DN15, TC |
| Branch process lines | 2 K Return | 3.8 | DN65 | 1 | - | 1 x DN65 |
| | 4.5 K Supply | 4.5 | DN15 | 1 | - | 1 x DN15 |
| | LTTS Return | 9 | DN15 | 1 | - | 1 x DN15 |
| | HTTS Supply | 40 | DN20 | 1 | - | 1 x DN20 |
| | HTTS Return | 80 | DN20 | 1 | - | 1 x DN20 |
| | CD Return from HTTS Return | 80 | DN20 | 1 | - | 1 x DN20 |
| | CD Return from 2 K Return | 80 | DN20 | 1 | - | 1 x DN20 |
| | CD Return from LTTS Return | 80 | DN20 | 1 | - | 1 x DN20 |

### 3.1. PIP II required Mass flow rates

Required mass flow rates of the LHe and GHe have been calculated based on static and dynamic heat loads at various temperature levels (2 K, 5 K, and 40 K) for the PIP-II Linac and static heat leak to the CDS. In this calculation, a total of 23 cryomodules for baseline Linac and 25 cryomodules for upgraded Linac have been taken into consideration. The 2 K heat exchanger in each cryomodule is taken with an effectiveness of 0.8. Required mass flow rates are summarized in Table 2.

Table 2. Required mass flow rates for PIP-II Linac for steady operation.

| Flow Rates [g/s] | CDS Transfer-Line | 4 (6)* HB650 | 9 LB650 | 7 SSR2 | 2 SSR1 | 1 HWR | Linac mass flow g/s |
|---|---|---|---|---|---|---|---|
| 4.5 K Supply | 2.0 | 10.8 | 6.0 | 5.1 | 4.4 | 7.5 | 151 (173)* |
| 2 K Return | 1.0 | 9.7 | 5.3 | 4.4 | 2.2 | 3.3 | 125 (145)* |
| LTTS Return | 1.0 | 1.1 | 0.8 | 0.8 | 2.2 | 4.2 | 26 (28)* |
| HTTS line | 1.0 | 2.0 | 1.5 | 2.2 | 3.3 | 3.1 | 48 (52)* |

( )* For Upgraded PIP-II Linac

### 3.2. Mass flow capacity of CDS for steady state operation

CDS steady state operation mass flow rate capacity has been evaluated for CDS process lines based on pressure drop budget based on anticipated Cryoplant operating requirements. Sizes of CDS lines are as summarized in Table 1. Result of this analysis is presented in Table 3.

**Table 3.** CDS mass flow capacity and pressure drop budget.

| Circuit | PIP-II Linac | |
|---|---|---|
| | Mass flow capacity [g/s] | Pressure drop budget [mbar] |
| 2 K Return | 146 | 4.3 |
| 4.5 K Supply | 178 | 30 |
| LTTS Return | 32 | |
| HTTS Supply and Return | 64 | 280 |

## 4. Design of the CDS for transient Linac operation

In addition to steady state operation, the CDS is also required to support transient operation of the Linac. The transient operation includes controlled cooldown/warm-up of the cryomodule thermal shield as well as slow cooldown, intermediate warm-up, and fast cooldown of the cryomodule cavities. The CDS design for supporting these modes is described in this section.

### 4.1. Cooldown and warmup of CDS

CDS cooldown flow capacity is estimated for 4.5 K circuit and high temperature thermal shield (HTTS) circuits. The 4.5 K circuit is divided into two loops: 4.5 K supply low temperature thermal shield (LTTS) loop-1 and 4.5 K supply 2 K return loop-2. The HTTS circuit is also divided into two loops: HTTS supply and return loop-1 and HTTS supply and cooldown return loop-2. Under the constraint on return pressure of the Cryoplant compressor suction of 1.5 bara, mass flow capacity of CDS is calculated for CDS line size and control valve size are given in Table 1. The estimated CDS flow capacity from room temperature to operating temperature is shown in Figure 3.

### 4.2. PIP-II Linac cool down and warm up

The 650 MHz cryomodule cool down will be done in three steps: initial cool down from 300 K to 5 K, intermediate warm-up to 45 K and soak, and fast cool down from 45 K to 4.5 K. Flow capacities for CDS and cryomodules are estimated for cool down with the constraint on return pressure of compressor of 1.5 bara and cavity warm MAWP of 2.05 bara. Results are shown in Figure 4. Warmup flow capacity calculations for the PIP-II Linac are shown in Figure 5.

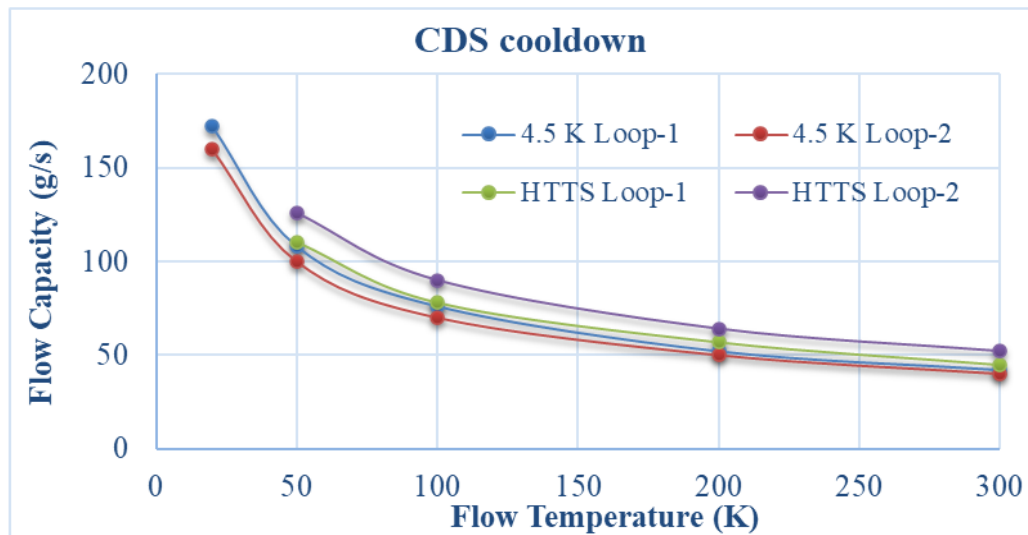

**Figure 3.** CDS cool down flow rate capacities vs. temperature.

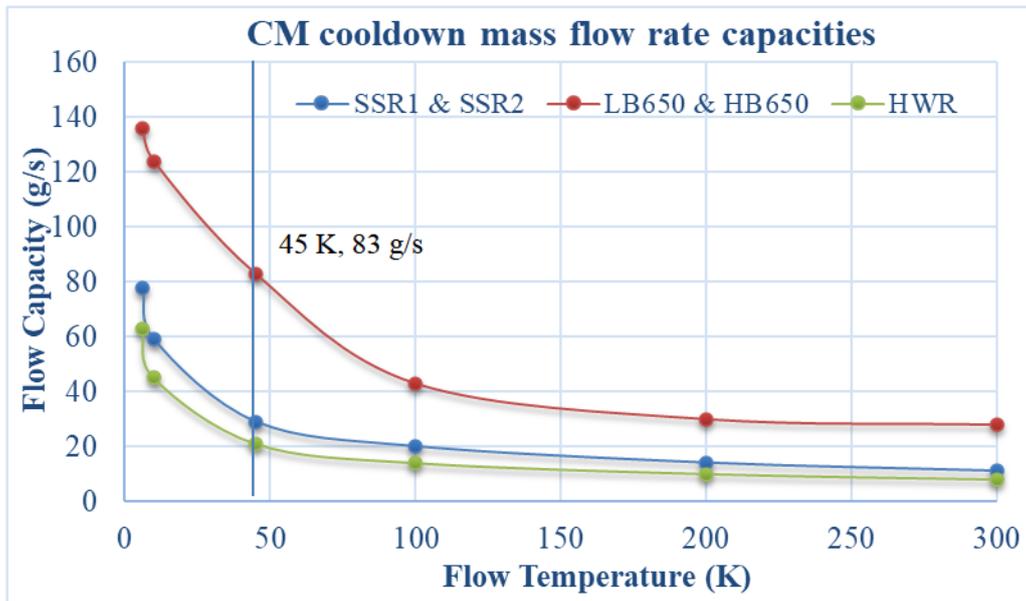

**Figure 4.** CM Cavity circuit cooldown flow capacity vs. temperature.

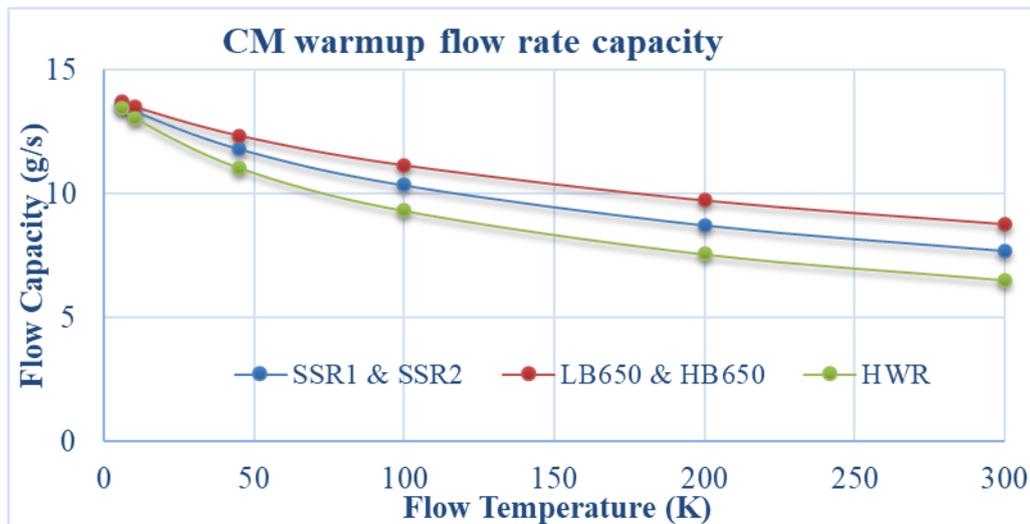

**Figure 5.** CM Cavity circuit warmup flow capacity vs. temperature.

### 4.3. Intermediate warmup of the CMs

The intermediate warmup of the CMs is assumed at a temperature up to 45 K. Mass flow rate capacity at this temperature has been estimated and summarized in Table 4. To achieve this mass flow rate at 45 K, one warm and one cold stream has been mixed.

**Table 4.** Flow rate capacity during intermediate warmup for CM cavity circuit at 45 K.

| Stream | Pressure [bara] | Temperature [K] | Mass flow rate [g/s] |
|---|---|---|---|
| Hot | 3.5 | 300 | 4 |
| Cold | 3 | 4.5 | 21 |
| Mixture | 3 | 45 | 25 |

## 5. Conclusion

The required mass flow rate during steady state operation for the upgraded Linac as summarized in Table 2 is compared with the CDS mass flow rate capacity summarized in Table 3 and it is found that CDS has sufficient mass flow over-capacity to run the upgraded Linac in steady state operation mode.

From the Linac cooldown analysis it is found that the during fast cooldown for HB650 and LB650 CDS can provide 83 g/s helium gas at 45 K.

**Acknowledgement**